\documentclass[preprint,12pt,3p,review]{elsarticle}

\usepackage{amssymb}
\usepackage{graphicx,amsmath}
\usepackage{natbib,lineno}
\biboptions{numbers,sort&compress}

\usepackage[breaklinks,colorlinks=true,citecolor={blue}, linkcolor= {blue}, urlcolor ={blue}]{hyperref}
\journal{Elsevier}

\begin{document}
\begin{frontmatter}
\title{Effect of screening on phonon-drag-induced Seebeck coefficient in bilayer graphene/AlGaAs quasi-2D electron gas structures}


\author[addr2,addr3]{Vo Van Tai}
\address[addr2]{Laboratory of Applied Physics, Science and Technology Advanced Institute, Van Lang University, Ho Chi Minh City, Vietnam}
\address[addr3]{Faculty of Applied Technology, School of Technology, Van Lang University, Ho Chi Minh City, Vietnam}

\author[addr2,addr3]{Nguyen Duy Vy\corref{cor1}}
\cortext[cor1]{Corresponding author email: nguyenduyvy@vlu.edu.vn}

\author[addr1]{Truong Van Tuan}
\address[addr1]{University of Tran Dai Nghia, 189-Nguyen Oanh Street, Go Vap District, Ho Chi Minh City, Vietnam}

\author[addr4]{Nguyen Quoc Khanh}
\address[addr4]{University of Science, VNU-HCM, Ho Chi Minh City, Vietnam}


\begin{abstract}
This study examines the temperature-dependent screening effect on the phonon-drag-induced Seebeck coefficient (Sg) in a bilayer graphene (BLG)-AlGaAs/quasi-two-dimensional electron gas (q2DEG) system. The BLG layer interacts with both deformation potential acoustic phonons and stronger piezoelectric field acoustic phonons from AlGaAs/GaAs. We compare the electron–phonon interactions in BLG with and without screening by q2DEG. The screening effect reduces Sg, particularly at low temperatures, and shows a strong dependence on the carrier density in the BLG layer. The double-layer screening function increases Sg with layer separation (d), paralleling the monolayer screening at large d. Additionally, varying the GaAs quantum well width (L) reveals that Sg increases with L < 100 Å under double-layer screening but remains unchanged beyond this threshold, while monolayer screening decreases Sg as L increases. Both screening functions enhance Sg when the BLG carrier density is lower than that of q2DEG, though the magnitude difference between them is minimal.
\end{abstract}


\end{frontmatter}

\section{Introduction}
The thermoelectric properties of low-dimensional systems have recently
garnered significant interest for applications in electronic devices \cite{Sanam, Xiao2022}. The screening effect in bilayer graphene (BLG)-quasi-two-dimensional electron gas (q2DEG) double layers, where
q2DEG is a GaAs semiconductor, has been previously studied \cite{Tuan2024}. When
the system is in air, the 2D electrons in BLG interact primarily with
deformation potential acoustic phonons. We have analyzed how the
screening effect from the second layer influences electron-phonon
interactions in the other layer and compared these results with those
using only monolayer screening functions. Introducing a semiconductor,
such as AlGaAs, between the BLG and q2DEG layers adds interactions with
intralayer piezoelectric field acoustic phonons from the semiconductor.

Key findings for the BLG-q2DEG system in air include: (1) Monolayer
screening functions in BLG and q2DEG are independent of the interlayer
distance d, with the double-layer screening function converging to the
monolayer functions at large d; (2) The screening effect reduces the
Seebeck coefficient (S\textsuperscript{g}\hspace{0pt}), particularly at
low temperatures, compared to monolayer screening alone; (3)
S\textsuperscript{g}\hspace{0pt} is strongly dependent on BLG when
carrier densities differ between the layers
(NsBLG$\neq$Nsq2DEGN\_s\^{}\{BLG\} $\neq$
N\_s\^{}\textsuperscript{q2DEG}NsBLG\hspace{0pt}=Nsq2DEG\hspace{0pt});
and (4) Varying the GaAs quantum well width L shows that the
double-layer screening function increases
S\textsuperscript{g}\hspace{0pt} for small L and has negligible effects
at large L, whereas monolayer screening decreases
S\textsuperscript{g}\hspace{0pt} as L increases.

When AlGaAs is introduced as the medium between BLG and q2DEG, the
screening effect on the phonon-drag-induced Seebeck coefficient exhibits
significant changes. In this paper, we investigate these effects in the
BLG-AlGaAs/q2DEG system. We examine how S\textsuperscript{g}\hspace{0pt}
is influenced by the screening effect from the second layer, comparing
it to the monolayer screening and to the BLG-q2DEG system in air. We
also explore the effects of temperature, carrier densities, interlayer
distance d, and the GaAs quantum well width L on
S\textsuperscript{g}\hspace{0pt}.

\section{Calculation methods}

The BLG-AlGaAs/q2DEG double layer is a structure consisting of a BLG
layer parallel to a q2DEG layer with width \emph{L} (here GaAs) placed a
distance \emph{d} apart from the base dielectrics as shown in Fig. 1.

The phonon-drag thermopower \emph{S}\textsuperscript{g} for the double
layer system reads \cite{Xiao2022,Smith2003},
where \(\text{$\sigma$}\) is the electrical conductivity given by
\(\text{$\sigma$} = N_{s}e^{2}\tau_{t}(E_{\mathbf{k}})/m^{*}\);
\(\text{u}\text{,}\text{\ l}\) correspond to upper and lower layers.

\begin{figure}[!h] 	\centering
\includegraphics[width=.9\textwidth]{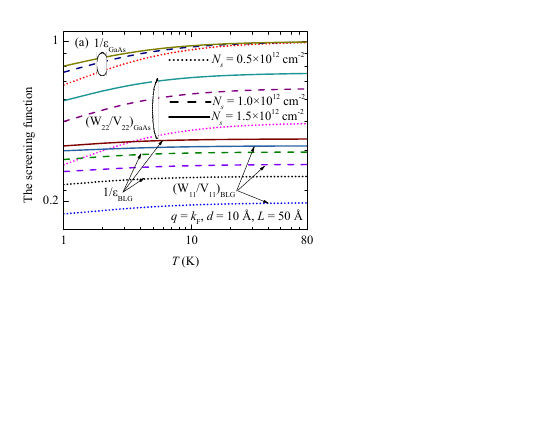}
\caption{The double layer structure consists of BLG and GaAs
quantum well, separated and encapsulated by different dielectric layers,
where \emph{$\kappa$}\textsubscript{2} is AlGaAs.} \label{fig_E2_metals}
\end{figure}

\textbf{Figure 1.} The double layer structure consists of BLG and GaAs
quantum well, separated and encapsulated by different dielectric layers,
where \emph{$\kappa$}\textsubscript{2} is AlGaAs.

In this work, we focus on the BLG-AlGaAs/q2DEG model with dielectric
substrates \emph{$\kappa$}\textsubscript{3} = 1, \emph{$\kappa$}\textsubscript{1} = 1,
\emph{$\kappa$}\textsubscript{2} = \emph{$\kappa$}\textsubscript{AlGaAs}. In this
scenario, the Seebeck phonon drag coefficient of the double-layer system
comprises the contributions from the deformation potential and the
piezoelectric field acoustic phonons with the following relations
\cite{Ansari2017, Ansari2021}:
\[\text{S}_{\text{DP}}^{\text{g}} = - \frac{m^{*3/2}D^{2}\mathcal{l}_{p}}{2\sqrt{2}N_{s}ek_{B}T^{2}\rho\pi^{2}\hslash^{3}v_{s}^{3}}\int_{0}^{\infty}{dq\ }\ \ \ \ \ \ \ \ \ \ \ \ \ \ \ \ \ \ \ \ \ \ \ \ \ \ \ \ \ \ \ \ \ \ \ \ \ \ \ \ \ \ \ \ \ \ \ \ \ \ \ \ \ \ \ \ \]

\[\times \left( \hslash\omega_{\mathbf{q}} \right)^{3}\int_{\gamma}^{\infty}{dE_{\mathbf{k}}G\left( E_{\mathbf{k}},\omega_{\mathbf{q}} \right)\left( \frac{W_{ii}(q,T)}{V_{ii}(q)} \right)^{2}N_{\mathbf{q}}\frac{\text{f}_{0}(E_{\mathbf{k}})\left\lbrack 1 - \text{f}_{0}\left( E_{\mathbf{k}} + \hslash\omega_{\mathbf{q}} \right) \right\rbrack}{\sqrt{E_{\mathbf{k}} - \gamma}}}\ \ \ \ \text{(2)}\]

\[\text{S}_{\text{PE}}^{\text{g}} = - \frac{m^{*3/2}C_{PE}^{2}D_{PE}^{2}\mathcal{l}_{p}}{8\sqrt{2}N_{s}ek_{B}T^{2}\rho_{GaAs}\pi^{3}\hslash^{3}v_{PE}^{2}}\int_{0}^{\infty}{dq}\ \ \ \ \ \ \ \ \ \ \ \ \ \ \ \ \ \ \ \ \ \ \ \ \ \ \ \ \ \ \ \ \ \ \ \ \ \ \ \ \ \ \ \ \ \ \ \ \ \ \ \ \ \ \ \ \ \ \ \ \ \ \ \ \ \ \ \ \ \ \ \ \ \ \ \ \ \ \ \ \ \ \ \ \ \]

\[\times \left( \hslash\omega_{PE} \right)^{2}\int_{\gamma}^{\infty}{dE_{\mathbf{k}}G\left( E_{\mathbf{k}},\omega_{\mathbf{q}} \right)\left( \frac{W_{ii}(q,T)}{V_{ii}(q)} \right)^{2}N_{\mathbf{q}}\frac{\text{f}_{0}(E_{\mathbf{k}})\left\lbrack 1 - \text{f}_{0}\left( E_{\mathbf{k}} + \hslash\omega_{\mathbf{q}} \right) \right\rbrack}{\sqrt{E_{\mathbf{k}} - \gamma}}}\ \ (\text{3})\]

where \(W_{ii}(q,T)\) is the double layer screening potential \cite{Scharf2012, Radi2013}

\[W_{ii}(q,T) = \frac{V_{ii}(q) + \left\lbrack V_{ii}(q)V_{jj}(q) - V_{ij}^{2}(q) \right\rbrack\Pi_{jj}(q,T)}{\left\lbrack 1 + V_{ii}(q)\Pi_{ii}(q,T) \right\rbrack\left\lbrack 1 + V_{jj}(q)\Pi_{jj}(q,T) \right\rbrack - V_{ij}^{2}(q)\Pi_{ii}(q,T)\Pi_{jj}(q,T)}\ \text{(4)}\]

and \(V_{ij}(q)\) is the Coulomb potential,

\[V_{ij}(q) = \frac{2\pi e^{2}}{q}\text{f}_{ij}(q)\ \ \ \ \ \ \ \ \ \ \ \ \ \ \ \ \ \ \ \ \ \ \ \ \ \ \ \ \ \ \ \ \ \ \ \ \ \ \ \ \ \ \ \ \ \ \ \ \ \ \ \ \ \ \ \ \ \ \ \ \ \ \ \ \ \ \text{(5)}\]

with \(\Pi_{ii}(q,T)\) being the polarization function at \emph{T} for
the \emph{i}-th layer with \emph{i}, \emph{j} =
\(\text{u},\text{\ }\text{l}\); given by expressions (12), (13). For a
double layer structure in figure 1, \(\text{u}\) = BLG, \(\ \text{l}\) =
GaAs, and\({\text{\ }\text{f}}_{ii}(q)\) have the form
{[}\protect\hyperlink{tai_lieu_1}{3},\protect\hyperlink{tai_lieu_6}{8}{]},

\[\text{f}_{jj}(x,y) = \frac{\kappa_{1}\kappa_{2}\left\lbrack \kappa_{2}\sinh{(x)} + \kappa_{3}\cosh{(x)} \right\rbrack}{\kappa_{2D}y^{2}\left( y^{2} + 4\pi^{2} \right)^{2}N(x,y)} \times \left\{ 64\pi^{4}\left\lbrack 1 - \cosh{(y)} \right\rbrack + y\left( y^{2} + 4\pi^{2} \right)\left( 3y^{2} + 8\pi^{2} \right)\sinh{(y)} \right\} + \frac{\left\lbrack \kappa_{2}\left( \kappa_{1} + \kappa_{3} \right)\cosh{(x)} + \left( \kappa_{2}^{2} + \kappa_{1}\kappa_{3} \right)\sinh{(x)} \right\rbrack}{y^{2}\left( y^{2} + 4\pi^{2} \right)^{2}N(x,y)} \times \left\lbrack y\left( 32\pi^{4} + 20\pi^{2}y^{2} + {3y}^{4} \right)\cosh{(y)} - 32\pi^{4}\sinh{(y)} \right\rbrack + \frac{\kappa_{2D}\left\lbrack \kappa_{2}\cosh{(x)} + \kappa_{3}\sinh{(x)} \right\rbrack y\left( y^{2} + 4\pi^{2} \right)\left( 3y^{2} + 8\pi^{2} \right)\sinh{(y)}}{y^{2}\left( y^{2} + 4\pi^{2} \right)^{2}N(x,y)}\ \ \ \ \ \ \ \ \ \ \ \ \ \ \ \ \ \ \ \ \ \ \ \ \ \ \ \ \ \ \ \ \ \ \text{\ }\text{\ \ \ \ }\text{(}\text{6}\text{)}\]

\(\text{f}_{ii}(x,y)\)

\[= \frac{2\kappa_{2}\cosh{(x)\left\lbrack \kappa_{1}\sinh{(y)} + \kappa_{2D}\cosh{(y)} \right\rbrack + 2\kappa_{2D}\sinh{(x)\left\lbrack \kappa_{1}\cosh{(y)} + \kappa_{2D}\sinh{(y)} \right\rbrack}}}{N(x,y)}\ \text{(}\text{7}\text{)}\]

\[\text{f}_{ij}(x,y) = \frac{8\pi^{2}\kappa_{2}\left\{ \kappa_{1}\left\lbrack \cosh{(y)} - 1 \right\rbrack + {\kappa_{2D}\sinh}{(y)} \right\}}{y\left( y^{2} + 4\pi^{2} \right)N(x,y)}\ \ \ \ \ \ \ \ \ \ \ \ \ \ \ \ \ \ \ \ \ \ \ \ \ \ \ \ \ \ \ \ \ \ \ \ \ \ \ \ \ \ \ \ \ \ \ \ \ \ \ \ \ \ \ \ \ \ \ \ \ \ \ \ \ \ \text{(}\text{8}\text{)}\]

with \(x = qd\), \(y = qL\), and \(N(x,y)\) given by

\[N(x,y) = \kappa_{2}\cosh{(x)}\left\lbrack \kappa_{2D}\left( \kappa_{1} + \kappa_{3} \right)\cosh{(y) + \left( \kappa_{1}\kappa_{3} + \kappa_{2D}^{2} \right)\sinh{(y)}} \right\rbrack\ \ \ \ \ \ \ \ \ \ \ \ \ \ \ \ \ \ \ \ \ \ \ \ \ \ \]

\[+ \sinh{(x)}\left\lbrack \kappa_{2D}(\kappa_{1}\kappa_{3} + \kappa_{2}^{2}{)cosh}{(y) + \left( \kappa_{1}\kappa_{2}^{2} + \kappa_{3}\kappa_{2D}^{2} \right)\sinh{(y)}} \right\rbrack\ \ \ \ \ \ \text{\ \ }\text{(}\text{9}\text{)}\]

The phonon-drag thermopower \emph{S}\textsuperscript{g} in q2DEG is
given by
\cite{Ando1982, Tuan2023}:

\[\text{S}_{\text{q2DEG}}^{\text{g}} = \sum_{\lambda = l,t}^{}\frac{\mathcal{l}_{p}m^{*\frac{3}{2}}v_{\lambda}}{16\sqrt{2}N_{s}ek_{B}T^{2}\rho\pi^{3}}\int_{0}^{\infty}{dq}\int_{0}^{\infty}{dq_{z}}q^{2}Q^{2}\left| C\left( \mathbf{Q} \right) \right|^{2}\left| I\left( q_{z} \right) \right|^{2}\hslash\omega_{\mathbf{Q}}\left( \frac{W_{jj}(q,T)}{V_{jj}(q)} \right)^{2}\]

\[\times \int_{\gamma}^{\infty}{dE_{\mathbf{k}}N_{\mathbf{Q}}\frac{\text{f}_{0}(E_{\mathbf{k}})\left\lbrack 1 - \text{f}_{0}\left( E_{\mathbf{k}} + \hslash\omega_{\mathbf{Q}} \right) \right\rbrack}{\sqrt{E_{\mathbf{k}} - \gamma}}}\ \ \ \ (10)\]

where \(\left| C(\mathbf{Q}) \right|^{2}\) is the intensity of
acoustical phonon-electron interactions consisting of the deformation
and piezoelectric potentials, \(\left| I(q_{z}) \right|^{2}\ \)is the
overlap integral in internal scattering, and \(\lambda = l,t\) for the
longitudinal and transverse acoustical phonons.

The dielectric function characterizing the screening effect in the
random phase approximation (RPA) reads
\cite{Sarma2011, Lv2010, Tuan2021Eur}:
\[\varepsilon(q,T) = 1 + \frac{2\pi e^{2}}{\overline{\kappa}q}\Pi(q,T)\ \ \ \ \ \ \ \ \ \ \ \ \ \ \ \ \ \ \ \ \ \ \ \ \ \ \ \ \ \ \ \ \ \ \ \ \ \ \ \ \ \ \ \ \ \ \ \ \ \ \ \ \ \ \ \ \ \ \ \text{(11)}\]

with \(\Pi(q,T)\ \)being the polarization function at the temperature
\emph{T}.

For BLG, the polarization function reads
\cite{Lv2010}:

\[\Pi_{\text{BLG}}(q,T) = \frac{\text{g}_{s}\text{g}_{\nu}m^{*}}{2\pi\hslash^{2}}\int_{0}^{\infty}\frac{dk}{k^{3}}\left\{ \sqrt{{4k}^{4} + q^{4}} \right.\  - k^{2} - \left| k^{2} - q^{2} \right| + \left\lbrack \text{f}(E_{\mathbf{k}}) + \text{f}(E_{\mathbf{k}} + 2\zeta_{\text{BLG}}) \right\rbrack\]

\[\left. \ \left\lbrack 2k^{2} - \sqrt{{4k}^{4} + q^{4}} + \frac{\left( 2k^{2} - q^{2} \right)^{2}}{q\sqrt{q^{2} - 4k^{2}}}\theta(q - 2k) \right\rbrack \right\}\ \ \ \ \ \ \ \ \ \ \ \ \ \text{(12)}\]

\[\Pi_{\text{q2DEG}}(q,T) = \int_{0}^{\infty}{d\mu'}\frac{\Pi(q,T = 0)}{4k_{B}T\cosh^{2}\left( \frac{\zeta_{\text{q2DEG}} - \mu^{'}}{2k_{B}T} \right)}\ \ \ \ \ \ \ \ \ \ \ \ \ \ \ \ \ \ \ \ \ \ \ \ \ \ \ \ \ \ \ \ \text{(13)}\]

where \(\Pi(q,T = 0)\) is the polarization function at 0 K,

\[\Pi(q,T = 0) = \frac{\text{g}_{s}\text{g}_{\nu}m^{*}}{2\pi\hslash^{2}}\left\lbrack 1 - \Theta\left( q - 2k_{F} \right)\sqrt{1 - \left( 2k_{F}/q \right)^{2}} \right\rbrack\text{,}\ \ \ \ \ \ \ \ \ \ \ \ \ \ \ \text{(14)}\]

In this case, the phonon-drag thermopower \emph{S}\textsuperscript{g}
for the \emph{i}-th layer has a similar expression to the one above but
with \(\frac{W_{ii}(q,T)}{V_{ii}(q)}\) being replaced by
\(\frac{1}{\varepsilon_{i}(q,T)}\), where \(\varepsilon_{i}(q,T)\) is
the monolayer screening potential.

\textbf{3. Results}

We investigate the phonon-drag thermopower \emph{S}\textsuperscript{g}
for the BLG-AlGaAs/q2DEG system with the following parameters. For BLG:
\emph{m\textsuperscript{*}} = 0.033\emph{m}\textsubscript{e}, \emph{D} =
20 eV, \(\rho =\) 7.6×10\textsuperscript{8} g/cm\textsuperscript{2},
v\textsubscript{s} = 2×10\textsuperscript{6} cm/s,
\(\mathcal{l}_{p}^{\text{BLG}} = \ \)10 µm,
\(\overline{\kappa} = \left( \text{1}\text{\ }\text{+}\text{\ }\text{$\kappa$}_{\text{2}} \right)\text{/2}\)
\cite{Kubakaddi2010}; \emph{D}\textsubscript{PE} =
2.4×10\textsuperscript{7}eV/cm, \emph{C}\textsubscript{PE} = 4.9,
v\emph{\textsubscript{PE}} = 2.7×10\textsuperscript{5} cm/s
\cite{Ansari2021}. For GaAs
\cite{VanTan2019,Sankes2005}:
\emph{m\textsuperscript{*} = m\textsubscript{z}} =
0.067\emph{m\textsubscript{e}}, \emph{$\rho$ =} 5.31 g/cm\textsuperscript{3},
\emph{u\textsubscript{l }} = 5.14×10\textsuperscript{5} cm/s,
\emph{u\textsubscript{t }} = 3.04×10\textsuperscript{5} cm/s,
\emph{D}\textsubscript{q2DEG} = 12 eV,
\(\overline{\kappa} = \kappa_{2D} = \text{12.91}\),
\emph{h\textsubscript{14}} = 1.2×10\textsuperscript{7} V/cm,
\(\mathcal{l}_{p}^{\text{q}\text{2DEG}} = \ \)0.1 mm,
\(\kappa_{\text{AlGaAs}} = \text{12.}\text{75}\).

In Fig. 2, we examine the behavior of the monolayer screening function
\(1/\varepsilon_{i}(q_{i},T)\), and the double layer screening function
\(W_{ii}/V_{ii}(q_{i},T)\) across three different conditions: (a)
variation with temperature at constant carrier densities in both layers,
(b) variation with carrier density at three different temperatures, and
(c) variation with the distance between the two layers at both equal and
differing carrier densities. Compared to the BLG-q2DEG system, we
observe that the screening functions in both the GaAs monolayer and
double layer are greater than those in the BLG monolayer and double
layer. However, in the BLG-AlGaAs/q2DEG system, the difference between
the two screening functions in the BLG layer is smaller than that in the
BLG-q2DEG system. For the GaAs system, Figs. 2(a) and 2(b) demonstrate
that both the monolayer and double layer screening functions increase
with temperature and carrier density before saturating at high values, a
behavior that parallels the BLG-q2DEG system. However, the difference
between the two screening functions at low densities is larger than that
observed in the BLG-q2DEG system. For the BLG system, the monolayer and
double layer screening functions remain nearly unchanged at high
temperatures, while both increase with carrier density, converging at
high densities. Figure 2(c) shows that the monolayer screening functions
of both BLG and GaAs are independent of the interlayer distance ddd,
with the double layer screening functions approaching the monolayer
values at large distances (\emph{d} \textgreater{} 100 Å) similar to the
BLG-q2DEG system. However, in the BLG-AlGaAs/q2DEG system, the double
layer screening functions increase with increasing distance \emph{d} for
both BLG and GaAs. In addition, with
\(\text{N}_{\text{s}}^{\text{BLG}}\ \)\textgreater{}
\(\text{N}_{\text{s}}^{\text{GaAs}}\)
(\(\text{N}_{\text{s}}^{\text{BLG}} <\)
\(\text{N}_{\text{s}}^{\text{GaAs}}\)) the double layer screening
functions of BLG (GaAs) coincide with those of the equal-density double
layer, independent of ddd at large distances, consistent with the
BLG-q2DEG system. It appears that \emph{$\kappa$}\textsubscript{2} has a
significant impact on both barrier functions.

\textbf{Figure 2.} Variation of the monolayer
1/$\epsilon$\emph{\textsubscript{i}}(\emph{q\textsubscript{i}, T}) and double
layer
\emph{W\textsubscript{ii}/V\textsubscript{ii}}(\emph{q\textsubscript{i},
T}), screening functions in the BLG-AlGaAs/q2DEG structure with respect
to \emph{T} (figure 2(a)), \emph{N}\textsubscript{s} (figure 2(b)), and
\emph{d} (figure 2(c)).

In Fig. 3, we show the change in \emph{S}\textsuperscript{g} with
temperature related to the deformation potential and the piezoelectric
field of acoustic phonons. The short line represents the screening
function \(W_{ii}/V_{ii}(q_{i},T)\), while the long line corresponds to
the screening function \(1/\varepsilon_{i}(q_{i},T)\). Figure 3(a)
illustrates how \emph{S}\textsuperscript{g} changes with the deformation
potential acoustic phonon, with the inset showing the BLG-q2DEG system.
We observe that \emph{S}\textsuperscript{g} increasing with temperature
is inversely proportional to the carrier density when using the
monolayer screening function (the double layer screening function) and
when \emph{T} \textless{} 20 K (\emph{T} \textless{} 30 K),
respectively; for the otherwise temperature regimes, it is proportional
to the carrier density for both the monolayer and double layer screening
functions. In comparison to the BLG-q2DEG system (the image is inserted
in Fig. 3(a)), \emph{S}\textsuperscript{g} is significantly larger in
magnitude when the system is BLG-AlGaAs/q2DEG. Figure 3(b) depicts the
variation of \emph{S}\textsuperscript{g} with the piezoelectric field
acoustic phonon, the inset corresponds to the total acoustic phonon
consisting of the piezoelectric field and deformation potential. We see
that \emph{S}\textsuperscript{g} increases with increasing temperature
and almost saturates at high temperatures. The total acoustic phonon is
mainly contributed by the PE scattering as pointed out by Ansari
\cite{Ansari2021} and the inset in Fig. 3(b).

\textbf{Figure 3.} (a) \emph{S}\textsuperscript{g} varies as a function
of \emph{T} at three carrier densities \emph{N\textsubscript{s}} =
0.5×1012
\emph{N\textsubscript{s}} = 1.0×10\textsuperscript{12}
cm$^{-2}$ and \emph{N\textsubscript{s}} =
1.5×10\textsuperscript{12} cm\textsuperscript{-2} of the deformation
potential acoustic phonon; the inserted image is the BLG-q2DEG system.
(b) \emph{S}\textsuperscript{g} changes as a function of \emph{T} of the
piezoelectric field acoustic phonon; the inserted image is the total
acoustic phonon including the piezoelectric field and deformation
potential.

\textbf{Figure 4.} The total phonon-drag coefficient,
\emph{S}\textsuperscript{g}, changes depending on the carrier density,
\emph{N}\textsubscript{s}, at \emph{T} = 5 K, \emph{T} = 10 K and
\emph{T} = 50 when using \(1/\varepsilon_{i}(q_{i},T)\) (dotted) and
\(W_{ii}/V_{ii}\left( q_{i},T \right)\ \)(solid). Figures 4(a) and 4(b)
show the results for equal and different carrier densities in the two
layers, respectively.

Figure 4 indicates the variation of \emph{S}\textsuperscript{g} with the
equal {[}Fig. 4(a){]} and different
{[}\(\text{N}_{\text{s}}^{\text{BLG}} = 10 \times \text{N}_{\text{s}}^{\text{GaAs}})\)
and vice versa, Fig. 4(b){]} carrier densities when using
\(1/\varepsilon_{i}(q_{i},T)\) (dotted) and
\emph{Wi}\(W_{ii}/V_{ii}(q_{i},T)\) (solid). From Fig. 4(a), we see that
\emph{S}\textsuperscript{g} decreases with the carrier density when
\emph{T} = 5 K, 10 K, and 50 K. For \emph{T} = 50 K,
\emph{S}\textsuperscript{g} with respect to the magnitude of the two
screening functions is almost identical regardless of densities
(compared to the BLG-q2DEG system, \emph{S}\textsuperscript{g} decreases
with the carrier density for two temperature values \emph{T} = 5 K, 10
K, while at \emph{T} = 50 K, \emph{S}\textsuperscript{g} increases with
the carrier density but quickly reaches saturation at high densities
{[}\protect\hyperlink{tai_lieu_1}{1}{]}). Moreover, when considering the
screening effect induced by the carriers in both the layers,
\emph{S}\textsuperscript{g} decreases compared to when using only the
monolayer screening effect, especially at low temperatures. Unlike the
BLG-q2DEG system, the separation between the values
\hspace{0pt}\hspace{0pt}of \emph{S\textsuperscript{g}} in the two
screening functions is negligible, especially at high temperatures. When
varying the density of the two layers (figure 4(b)), it is shown that
\emph{S}\textsuperscript{g} depends more strongly on the BLG layer than
on the BLG-q2DEG system.

\textbf{Figure 5.} \emph{S}\textsuperscript{g} of the double layer
system varies as a function of \emph{d} for
\(\text{N}_{\text{s}}^{\text{\ BLG}} = \text{N}_{\text{s}}^{\text{\ GaAs}} = \text{N}_{\text{s}}\text{\ =\ 1.5×}\text{10}^{\text{12}}\text{cm}^{\text{-2}}\),\(\ \text{N}_{\text{s}}^{\text{\ BLG}} = \text{N}_{\text{s}}\text{;}{\text{\ }\text{N}}_{\text{s}}^{\text{\ GaAs}} = \text{N}_{\text{s}}\text{/3}\),\(\ \)and
\(\text{N}_{\text{s}}^{\text{\ BLG}} = \text{N}_{\text{s}}\text{/3;}{\text{\ }\text{N}}_{\text{s}}^{\text{\ GaAs}} = \text{N}_{\text{s}}\),
where both \(1/\varepsilon_{i}(q_{i},T)\) và \(W_{ii}/V_{ii}(q_{i},T)\),
are utilized for separate calculations.

The graphical representation in Fig. 5 effectively demonstrates the
variation of \emph{S}\textsuperscript{g} as a function of \emph{d} using
both \(1/\varepsilon_{i}(q_{i},T)\) and \(W_{ii}/V_{ii}(q_{i},T)\),
while considering equal and different carrier densities in the two
layers for separate calculations. For the monolayer screening function,
\emph{S}\textsuperscript{g} does not depend on \emph{d} as already
indicated in figure 2(c). When using the double layer screening
function, \emph{S}\textsuperscript{g} increaes with \emph{d} and becomes
parallel with the previous case when taking the internal screening
function at large \emph{d} (\emph{d} \textgreater{} 100 Å).

When comparing the BLG-q2DEG system, we see that if we consider the
screening double layer function, \emph{S}\textsuperscript{g} decreases
with \emph{d} and is almost parallel to \emph{S}\textsuperscript{g} when
we consider the monolayer screening function at large \emph{d} (\emph{d}
\textgreater{} 50 Å). Therefore, the screening effect of the second
layer (i.e. GaAs) on the carrier-phonon interactions in the first layer
(i.e. BLG) becomes smaller at larger \emph{d}.

Variation of \emph{S}\textsuperscript{g} as a function of \emph{L} when
using both \(1/\varepsilon_{i}(q_{i},T)\) and
\(W_{ii}/V_{ii}\left( q_{i},T \right)\ \)with equal and different
carrier densities in the two layers for separate calculations is
illustrated in Fig. 6. When the quantum well width \emph{L} is less than
100 Å, the double layer screening function causes
\emph{S}\textsuperscript{g} to increase and then saturate for larger
quantum well widths (\emph{L} \textgreater{} 100 Å). This is different
from the behavior of the BLG-q2DEG system for \emph{L} \textless{} 50 Å
and \emph{L} \textgreater{} 50 Å. In the meantime, the monolayer
screening function induces \emph{S}\textsuperscript{g} to decrease with
\emph{L}, and both the monolayer and double layer screening functions
make \emph{S}\textsuperscript{g} greater for
\(\text{N}_{\text{s}}^{\text{\ BLG}}\) \textless{}
\({\text{\ }\text{N}}_{\text{s}}^{\text{\ GaAs}}\) similarly to the
BLG-q2DEG system. However, the difference in \emph{S}\textsuperscript{g}
magnitude between the two screening functions is negligible compared to
the BLG-q2DEG system.

\textbf{Figure 6.} \emph{S}\textsuperscript{g} of the double layer
system varies as a function of \emph{L} for
\(\text{N}_{\text{s}}^{\text{\ BLG}} = \text{N}_{\text{s}}^{\text{\ GaAs}} = \text{N}_{\text{s}}\text{\ =\ }\text{1.5×}\text{10}^{\text{12}}\text{cm}^{\text{-2}}\);
\(\text{N}_{\text{s}}^{\text{\ BLG}} = \text{N}_{\text{s}}\text{,}\text{\ N}_{\text{s}}^{\text{\ GaAs}} = \text{N}_{\text{s}}\text{/3}\);
and
\(\text{N}_{\text{s}}^{\text{\ BLG}} = \text{N}_{\text{s}}\text{/3,}\text{\ N}_{\text{s}}^{\text{\ GaAs}} = \text{N}_{\text{s}}\),
where both \(1/\varepsilon_{i}(q_{i},T)\) and \(W_{ii}/V_{ii}(q_{i},T)\)
are used for separate calculations.

\textbf{4. Summary and conclusions}

We conducted a detailed investigation into the phonon drag coefficient
S\textsuperscript{g}\hspace{0pt} of the BLG-AlGaAs/q2DEG system,
utilizing both monolayer and double layer screening functions. These
results were compared with those of the BLG-q2DEG system in an air
environment {[}3{]}. For BLG, the S\textsuperscript{g}\hspace{0pt} of
the total acoustic phonon is primarily influenced by the piezoelectric
field rather than the deformation potential, leading to a significantly
larger S\textsuperscript{g}\hspace{0pt}\hspace{0pt} in the
BLG-AlGaAs/q2DEG system compared to the BLG-q2DEG system. When
considering the monolayer and double layer screening functions
separately, the analyses revealed similar variations in the two
screening functions as observed in the BLG-q2DEG system. However, in the
BLG-AlGaAs/q2DEG system, the influence of $\kappa$\textsubscript{2}\hspace{0pt}
on the screening functions differed; specifically, the double layer
screening function increased with the distance d for both BLG and q2DEG.

For the BLG-AlGaAs/q2DEG system, the Seebeck coefficient
S\textsuperscript{g}\hspace{0pt} was examined in both symmetric and
asymmetric cases with equal and differing carrier densities in the two
layers. This investigation underscores the significant impact of the
second layer\textquotesingle s screening effect, such as the GaAs
quantum well, on electron-phonon interactions in the first layer (i.e.,
BLG). When both monolayer and double layer screening effects are
considered, S\textsuperscript{g}\hspace{0pt} increases at low
temperatures and saturates at higher temperatures. Furthermore, when
only the double layer screening effect is considered, the magnitude of
S\textsuperscript{g}\hspace{0pt} decreases compared to the monolayer
screening effect, particularly at low temperatures. When the carrier
densities in the two layers differ,
S\textsuperscript{g}\hspace{0pt}\hspace{0pt} becomes strongly dependent
on the first layer (BLG).

The impact of the distance d between the BLG and GaAs layers is as
follows: if only the monolayer screening function is considered,
S\textsuperscript{g}\hspace{0pt}\hspace{0pt} is independent of d; if the
double layer screening function is considered,
S\textsuperscript{g}\hspace{0pt} increases with increasing d and remains
almost parallel to S\textsuperscript{g}\hspace{0pt} under the monolayer
screening function at large d (in contrast to the BLG-q2DEG system,
where S\textsuperscript{g}\hspace{0pt}\hspace{0pt} decreases with d).
Finally, altering the GaAs quantum well width LLL shows that the double
layer screening function increases S\textsuperscript{g}\hspace{0pt} for
small LLL and remains nearly constant for large LLL, while the monolayer
screening function causes S\textsuperscript{g}\hspace{0pt} to decrease
with increasing LLL. Both the screening functions induce an increase in
\emph{S}\textsuperscript{g} when the density of
\(\text{N}_{\text{s}}^{\text{\ BLG}}\) \textless{}
\({\text{\ }\text{N}}_{\text{s}}^{\text{\ GaAs}}\). However, there is a
negligible difference in the magnitude of \emph{S}\textsuperscript{g}
between the two screening functions when compared with the BLG-q2DEG
system.


\end{document}